# Strain engineered direct-indirect band gap transition and its mechanism in 2D phosphorene

Xihong Peng, [1]* Andrew Copple, [2] Qun Wei[1, 3]

[1]School of Letters and Sciences, Arizona State University, Mesa, Arizona 85212, USA

[2]Department of Physics, Arizona State University, Tempe, Arizona 85287, USA

[3]School of Physics and Optoelectronic Engineering, Xidian University, Xi'an, 710071, P.R. China

**ABSTRACT**: Recently fabricated two dimensional (2D) phosphorene crystal structures have demonstrated great potential in applications of electronics. In this work, strain effect on the electronic band structure of phosphorene was studied using first principles methods. It was found that phosphorene can withstand a surface tension and tensile strain up to 10 N/m and 30%, respectively. The band gap of phosphorene experiences a direct-indirect-direct transition when axial strain is applied. A moderate -2% compression in the zigzag direction can trigger this gap transition. With sufficient expansion (+11.3%) or compression (-10.2% strains), the gap can be tuned from indirect to direct again. Five strain zones with distinct electronic band structure were identified and the critical strains for the zone boundaries were determined. The origin of the gap transition was revealed and a general mechanism was developed to explain energy shifts with strain according to the bond nature of near-band-edge electronic orbitals. Effective masses of carriers in the armchair direction are an order of magnitude smaller than that of the zigzag axis indicating the armchair direction is favored for carrier transport. In addition, the effective masses can be dramatically tuned by strain, in which its sharp jump/drop occurs at the zone boundaries of the direct-indirect gap transition.

Two dimensional (2D) layered crystal materials have attracted extensive research efforts in recent years, such as graphene [1, 2] and molybdenum disulfide [3], for their potential applications in future electronics. Most recently, researchers have successfully fabricated a new 2D few-layer black phosphorus [4-7] and found that this material is chemically inert and has great transport properties. It was reported that it has a carrier mobility up to 1000 cm$^2$/V·s [4] and an on/off ratio up to $10^4$ [5] was achieved for the phosphorene transistors at room temperature. In addition, this material shows a finite direct band gap at the center of Brillouin zone [4, 5, 8-10] (in contrast to the vanishing gap in graphene), which opens doors for additional applications in optoelectronics.

Tailoring electronic properties of semiconductor nanostructures has been critical for their applications. Strain has long been used to tune electronic properties of semiconductors [11, 12]. As a practical issue, strain is almost inevitable in fabricated monolayer nanostructures, manifesting as the formation of ridges and buckling [13, 14].  But a more interesting case comes from intentionally introduced and controlled strains. The approaches introducing strain include lattice mismatch, functional wrapping,[15, 16] material doping,[17, 18] and direct mechanical application.[19] It was found that nanostructures maintain integrity under a much higher strain than their bulk counterpart, [20, 21] which dramatically expands the strength of applicable strain to nanostructures. In particular, 2D layered materials, such as graphene and $MoS_2$, possess superior mechanical flexibility and can sustain a spectacularly large strain ($\geq$ 25%) [22-24]. In this work, we calculated the tension-strain relation in phosphorene and found that it can withstand a tensile strain up to 30%.

As already demonstrated by several research groups, strain shows remarkable effects in modifying the electronic properties of phosphorene. For example, Rodin *et. al.* [9], using density functional theory and tight-binding models, predicted an anisotropic dispersion relation with a direct band gap for phosphorene. They analyzed the localized orbital composition of the band edge states and suggested a semiconductor-to-metal transition with compression. Liu *et. al.* [5] briefly reported the sensitive dependence of the band gap on in-layer stress and a critical compressive strain of 3% to trigger the direct-to-indirect band gap transition. Fei and Yang [25] theoretically predicted that the preferred conducting direction in phosphorene can be rotated by 90 degrees with an appropriate biaxial or uniaxial strain based on the anisotropicity of the material [5, 9, 26].

However, a full picture of detailed and systematic analysis of the strain effect on the band structure is still missing. For example, what is the elastic limit for phosphorene? Where is the conduction/valence band edge located when the band gap becomes indirect? Is there any additional critical strain to trigger the band gap transition? What is the mechanism/origination for this gap transition? In present work, we will answer these questions by providing tension-strain relation and a full analysis of strain effects on the band structure of phosphorene. By revealing the evolution of band structure with strain, it is clear to see the shift of the band edges and gap transitions. Several critical strains have been identified to trigger the direct-indirect transition. In addition, with sufficient large tensile (+11.3%) or compressive (-10.2%) strain, the indirect band gap was found to become direct again. Five strain zones with distinct band structure were identified. We also discuss in detail the mechanism for the gap transition by examining the bond nature of near-band-edge electronic orbitals. This mechanism has been applied successfully in many other semiconductor nanostructures [27-36]. Effective masses of charge carriers (thus carrier mobility) were also found to be drastically tuned by strain.

**Calculation Details.** The first principles density functional theory (DFT) [37] calculations were carried out using the Perdew-Burke-Ernzerhof (PBE) exchange-correlation functional [38] along with the projector-augmented wave (PAW) potentials [39, 40] for the self-consistent total energy calculations and geometry optimization. The calculations were performed using the Vienna Ab-initio Simulation Package (VASP) [41, 42]. The kinetic energy cutoff for the plane wave basis set was chosen to be 350 eV. The reciprocal space was meshed at $14 \times 10 \times 1$ using Monkhorst-Pack method. The energy convergence criteria for electronic and ionic iterations were set to be $10^{-5}$ eV and $10^{-4}$ eV, respectively. Such parameter setting ensures the calculations were converged within 5 meV in total energy per atom. To simulate a monolayer of phosphorene, a unit cell with periodic boundary condition was used. A vacuum space of at least 16 Å was applied to minimize the interaction between layers. In band structure calculations, 21 points were collected along each high symmetry line in reciprocal space.

Unlike a flat structure of graphene, the single layer black phosphorus has a puckered honeycomb structure with each phosphorus atom covalently bonded with three adjacent atoms (see Figure 1). The initial structure of phosphorene was obtained from black phosphorus [43]. Our calculated lattice constants for bulk black phosphorus are $a = 3.308$ Å, $b = 4.536$ Å, and $c =$ 11.099 Å, in good agreement with experimental values [43] and other theoretical calculations [5, 26]. The relaxed lattice constants for a monolayer of phosphorene are $a = 3.298$ Å, $b = 4.627$ Å.

Starting with the relaxed phosphorene, strain within the range of ±12% was applied in either the x (zigzag) or y (armchair) direction to explore its effects on the band structure. The applied strain is defined as $\varepsilon = \dfrac{a - a_0}{a_0}$, where $a$ and $a_0$ are the lattice constants of the strained and relaxed structure, respectively. The positive values of strain refer to expansion, while negative corresponds to compression. With each axial strain applied, the lattice constant in the transverse direction was fully relaxed through the technique of energy minimization to ensure the force in the transverse direction is a minimum.

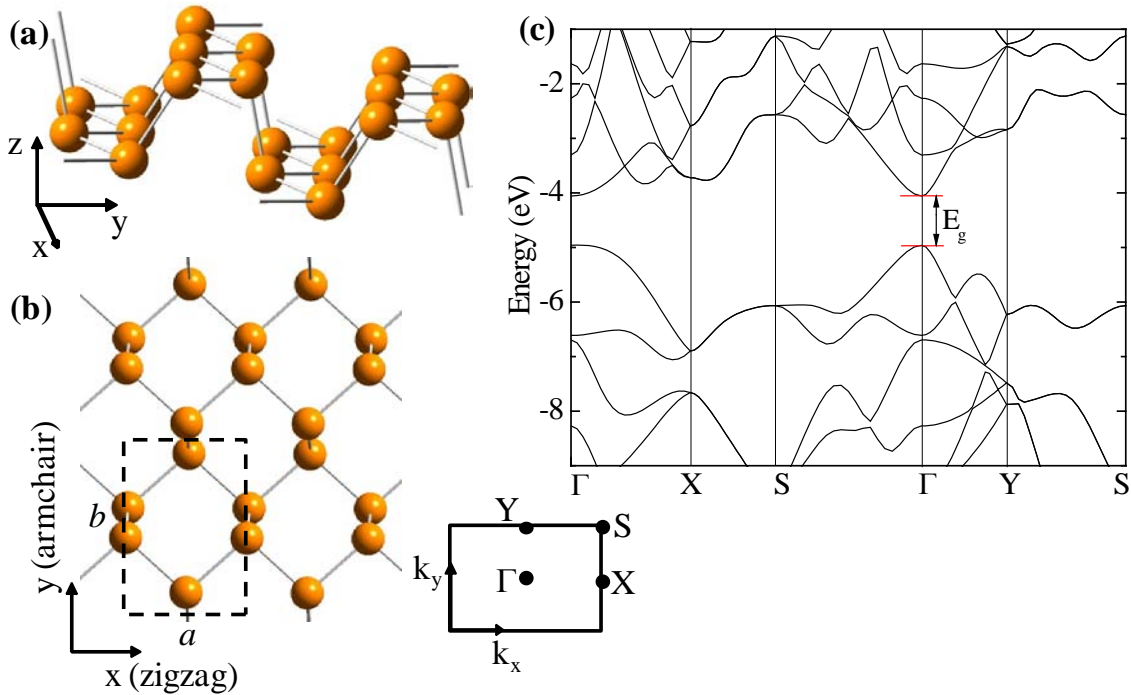

**Figure 1. Snapshots of 2D phosphorene and its band structure. The dashed rectangle in (b) indicates a unit cell. The energy in the band structure is referenced to vacuum level.**

**Tension-Strain Relation and Critical Strain Limit of Phosphorene.** Generally, a compression applied in an axial direction results an expansion in the transverse direction. The applied axial strains in the x (y) direction and the transverse strain response in the y (x) direction are reported in Figure 2(a). The Poisson's ratio, $\nu = -\dfrac{d\varepsilon_{transverse}}{d\varepsilon_{axial}}$, was found to be 0.7 and 0.2 for the axial strain applied in the zigzag and armchair directions, respectively. These significantly different Poisson's ratios further indicate the anisotropic nature of phosphorene. Figure 2(b) presents the change in the total energy of phosphorene as a function as the applied axial strain. The deeper energy surface for $\varepsilon_x$ suggests that strain is more difficult to apply in the zigzag than the armchair direction.

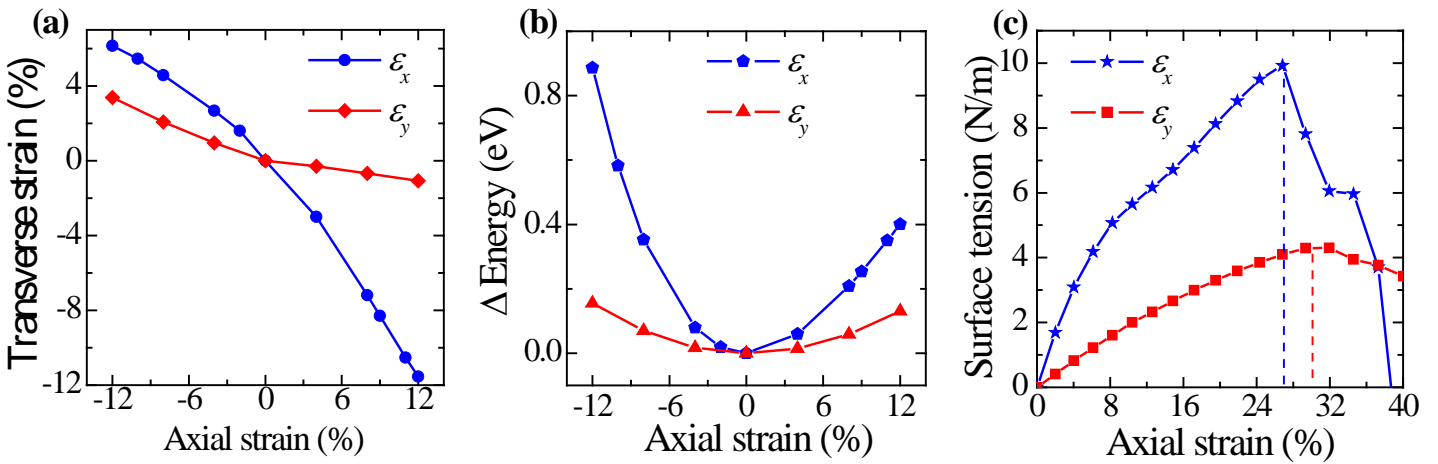

**Figure 2. (a) The applied axial strains in the x and y axes and their transverse strain response in the y and x directions, respectively. (b) The change in the total energy as a function of the applied strain. (c) The surface tension as a function of tensile load for phosphorene. Phosphorene can withstand a critical tensile strain up to 30% in the armchair direction. The different behaviors along the x and y directions in (a)-(c) indicate the anisotropic nature of phosphorene.**

To estimate the elastic limit of phosphorene, we calculated the surface tension (force per unit length) [44] of phosphorene as a functional of tensile strain, using the method described in the references [45, 46]. This method of calculating stress-strain relation was originally introduced for three dimensional crystals. In a 2D monolayer phosphorene, the stress calculated from the Hellmann-Feynman theorem was modified to be the surface tension [44].

The tensile strain is loaded in either the zigzag or armchair direction. Our calculated tension-strain relation is presented in Figure 2(c). It shows that phosphorene can sustain a surface tension up to 10 N/m and 4 N/m in the zigzag and armchair directions, respectively. The corresponding tensile strain limits are 27% and 30% along the zigzag and armchair axes, respectively. This predicted elastic strain limit is close to that found in other 2D materials such as graphene and $MoS_2$ [22-24], suggesting that phosphorene is highly flexible and may have potential applications in flexible display.

**Strain Effect on the Band Structure of Phosphorene.** Our DFT predicted band gap for 2D phosphorene is 0.91 eV, which is in agreement with other theoretical work [5, 10]. It is well known that DFT underestimates the band gap of semiconductors, and advanced GW method can provide improved predictions. The GW gap for phosphorene calculated by Yang's group [10] is about 2.0 eV. However, the present work is mainly focused on the strain effect on the band structure and previous studies [28] on Si nanoclusters showed that the energy gap calculated by DFT obeys a similar strain-dependency as the optical gap predicted by the advanced configuration interaction method and the quasi-particle gap. Therefore, we expect that DFT can correctly predict the general trends of strain effect on the band structure and near-gap states in phosphorene.

By comparing the band structure of phosphorene at different strain, we found that strain has a remarkable effect on the band structure along two particular k directions, Γ to X (0.5, 0, 0) and Γ to Y (0, 0.5, 0), respectively. For simplification, we only plot the band structure along these two directions in Figures 3 and 4.

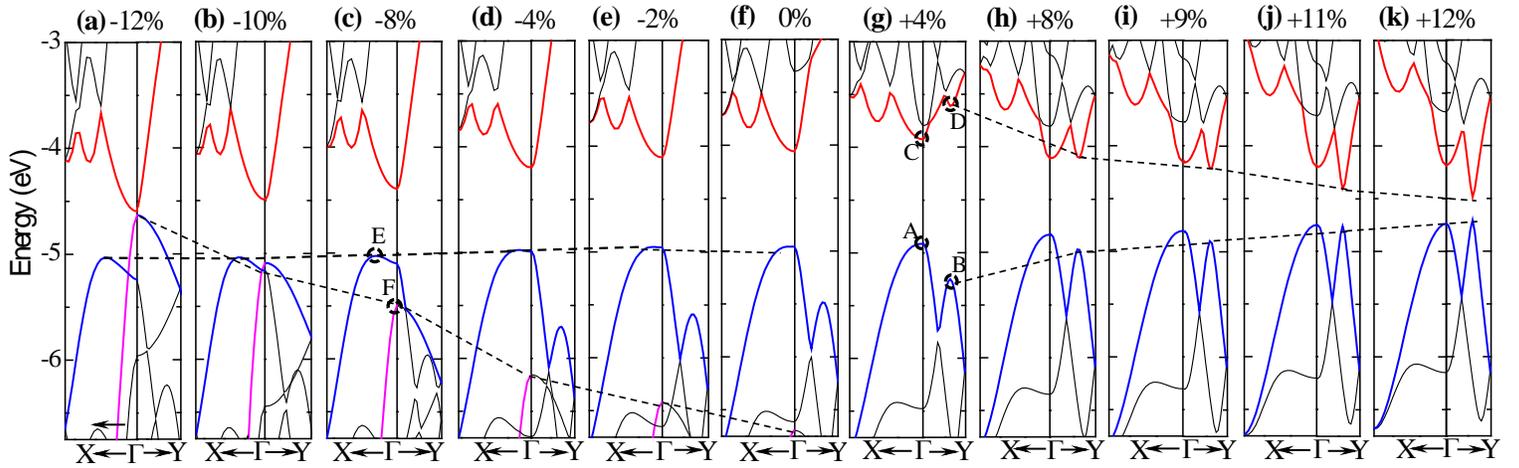

**Figure 3. The strain $\varepsilon_x$ (applied in the zigzag direction) manipulated direct-indirect band gap transition in 2D phosphorene. Positive strain means expansion while negative corresponds to compression. All energies are referenced to vacuum level. Starting from the relaxed structure, the band gap experiences a direct-indirect-direct transition with both tensile and compressive strain. The direct/indirect nature of the band gap is the result of the competition among the energies of near-band-edge states A – F. The dashed lines are guide for eye for the energy shifts of states B, D, E, F.**

Figure 3 presents the effect of strain $\varepsilon_x$ (applied in the zigzag direction) on the band structure. Figure 3(f) is for the relaxed phosphorene with a direct band gap at Γ. With an increase of tensile strain, the band gap becomes indirect, then back to direct. For example, at $\varepsilon_x$ = +9%, it shows an indirect band gap with the conduction band minimum (CBM) shifted from Γ to K1 (0, 0.3, 0) while the valence band maximum (VBM) remains at Γ. At $\varepsilon_x$ = +12%, it gives a direct gap with both the CBM and VBM at K1 (0, 0.3, 0). Similarly, with an increase in compression, the gap first transits to indirect and then back to direct. For instance, at $\varepsilon_x$ = -8%, the gap is indirect with the CBM remains at $\Gamma$ while the VBM shifted to K2 (0.15, 0, 0). When $\varepsilon_x$ = -12%, it shows a direct gap with both the CBM and VBM at $\Gamma$.

Figure 3 clearly demonstrates that the direct/indirect nature of the band gap is the result of the competition of the energies of several near-band-edge states. With an increase of tensile strain, the energy of the conduction band (CB) at K1 (0, 0.3, 0), labeled as state D in Figure 3(g), reduces rapidly and becomes equal to that of state C (CB at $\Gamma$, i.e. the original CBM) when $\varepsilon_x$ = +8%. With the strain higher than +8%, state D has a lower energy than C, thus D becomes the CBM. Similar behavior occurs with the valence band (VB). The energy of state B at K1 (0, 0.3, 0) increases faster than A (VB at $\Gamma$, i.e. the original VBM). At $\varepsilon_x$ = +12%, B has a higher energy than A, and thus represents the VBM.

On the side of compression, the CBM is always located at $\Gamma$ (state C). However, the VBM demonstrates an interesting transition. First, at $\varepsilon_x$ = -2%, the VB at K2 (0.15, 0, 0), labeled as state E in Figure 3(c), shows an equal energy with A (i.e. VB at $\Gamma$). With increased compression, the energy of state E is higher than that of A, indicating a direct-to-indirect gap transition. At $\varepsilon_x$ = -12%, the energy of a sub-VB (the pink band) state F at $\Gamma$ has a higher energy than E and becomes the VBM, showing a direct band gap at $\Gamma$.

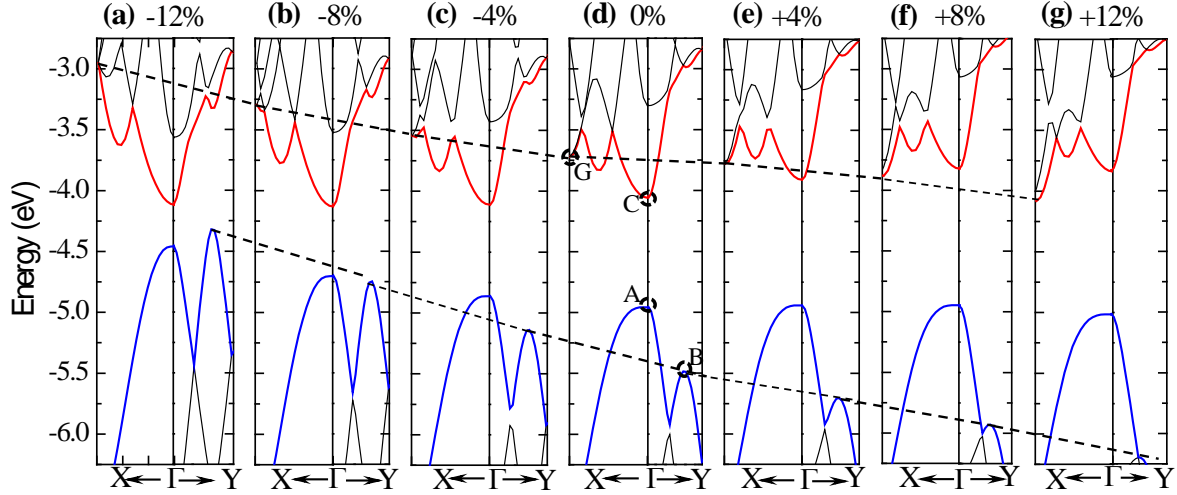

**Figure 4.** The strain $\varepsilon_y$ (applied in the armchair direction) manipulated direct-indirect band gap transition in phosphorene. All energies are referenced to vacuum level. The band gap shows a direct-indirect transition with expansion/compression. The direct/indirect nature of the gap results from the energy competition of near-band-edge states A, B, C, and G. The dashed lines are a guide for eye for the energy shifts of states B and G.

Figure 4 shows the effect of strain $\varepsilon_y$ (applied in the armchair direction) on the band structure of phosphorene. It demonstrates that both expansion and compression can convert the band gap to indirect. For instance, at $\varepsilon_y$ = +8% and +12%, the gap is indirect with the VBM at Γ while the CBM shifted from Γ to X. At the side of negative strain, the gap is indirect at $\varepsilon_y$ = -12% with the VBM at K1 (0, 0.3, 0). It is clear that two VB states A and B compete for the VBM, and two CB states C and G (CB at X) race for the CBM. The energy shift of states B and G with strain are more prominent than their competing states A and C, respectively.

**Strain Effect on the Band Gap of Phosphorene.** The band gap is defined as the energy difference between the CBM and VBM. From Figures 3 and 4, the gap strongly depends on strain. In Figure 5, the gap is plotted as a function of both strain $\varepsilon_x$ and $\varepsilon_y$.

For the strain applied along the zigzag direction in Figure 5(a), the band gap is initially increased with tensile strain from a value of 0.91 eV for the relaxed structure and reaches the maximal value of 0.99 eV at +4% strain, then drops rapidly with further increased expansion. At $\varepsilon_x = +12\%$, the band gap is reduced to 0.22 eV. To see if the gap reduces to zero with further increased tensile strain, we explored even larger strain +13% up to +16% with a 1% increment. The gap was not found to close. It reaches a minimal value of 0.06 eV at +14% strain, and then opens up again for larger strain.

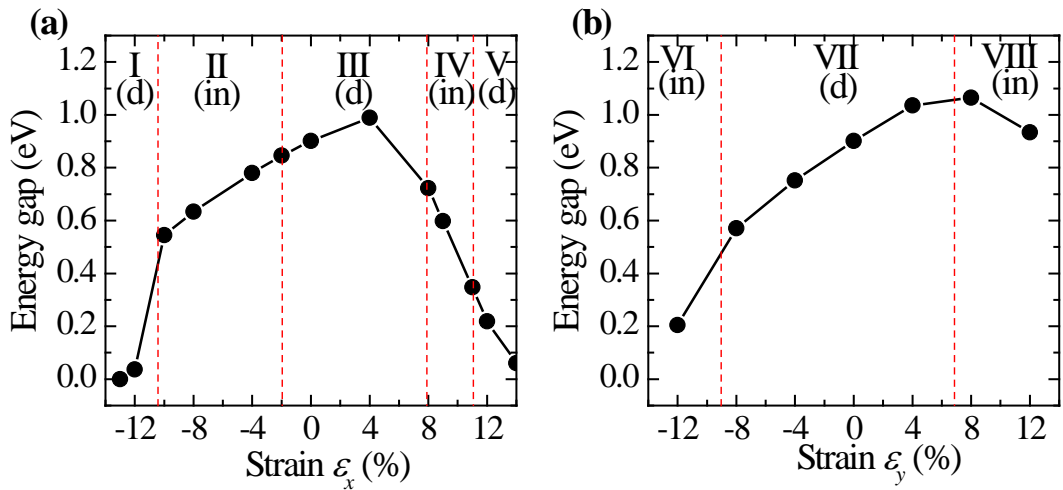

**Figure 5. The band gap of 2D phosphorene as a function of strain applied in the zigzag (left) and armchair (right) directions, respectively. Five (three) strain zones were identified for $\varepsilon_x$ ($\varepsilon_y$) based on its distinct band structures. Zones I, II, III, IV, and V in (a) are corresponding to the direct (d), indirect (in), direct, indirect, and direct gap, respectively. The critical strain for the gap transition are -10.2%, -2%, +8%, and +11.3%. The gap closes up at $\varepsilon_x$ = -13%. Zones VI, VII, and VIII in (b) present the indirect, direct, and indirect gap, respectively and the critical strains of the zone borders are $\varepsilon_y$ = -9% and +6.8%.**

On the side of negative strain $\varepsilon_x$, the gap reduces, mainly resulting from the downward shift of the CBM (see Figure 3). At $\varepsilon_x = -12\%$, the gap sharply drops to 0.04 eV from the value of 0.55 eV at -10% strain. This is due to the fact that the VBM was replaced by a newly raised sub-valence band (state F). To explore if the band gap reduces to zero with a larger compression, we explored -13% strain and find that the gap indeed closed up.

In Figure 5(a), five strain zones were identified based on their distinct band structure. Zone I is for a direct band gap within the strain range -12% to -10.2%, in which the CBM is represented by state C and the VBM is given by state F. Zone II corresponds to an indirect band gap from -10.2% to -2%, where the VBM is state E. Zone III is a direct gap at Γ from -2% to +8%. Zone IV is an indirect gap from +8% to +11.3%, where the CBM is at (0, 0.3, 0). Zone V shows a direct band gap with both the CBM and VBM at (0, 0.3, 0). The critical strains of -10.2%, -2%, +8% and +11.3% for the zone boundaries were determined in the next section.

Figure 5(b) presents the band gap as a function of strain applied in the armchair direction. The reduced gap value with compression is mainly resulted from the downward shift of the CBM and upward change of the VBM (see Figure 4). The drop of the gap value at +12% resulted from the fact that the CBM is replaced by the CB at X (state G). Three unique strain zones were characterized. Zone VI is for an indirect band gap from -12% to -9% where the VBM is located at (0, 0.3, 0). Zone VII is for direct band gap at Γ from -9% to +6.8%. Zone VIII shows an indirect band gap in the strain range +6.8% to +12%, in which the CBM is at X.

**Strain Effect on the Near-band-edge Orbitals.** To determine the critical strain in which the direct-indirect band gap transition occurs, we plotted the energies of the near-band-edge states A-G (labeled in Figure 3 and 4) as a function of strain in Figure 6. As seen from Figures 3 and 4, these are the states which compete for the CBM and VBM, thus their energies determine the direct/indirect nature of the gap. In Figure 6, the energies of the states display a nearly linear function with strain. For instance, the energy of state B increases with a positive strain $\varepsilon_x$ while decreases with a negative strain $\varepsilon_x$. In contrast, states D and F show an opposite trends. The energy of state E remains unaffected by strain. This linear shift of energy is not unique to phosphorene. It is also observed in other semiconducting nanostructures [27-36, 47-50].

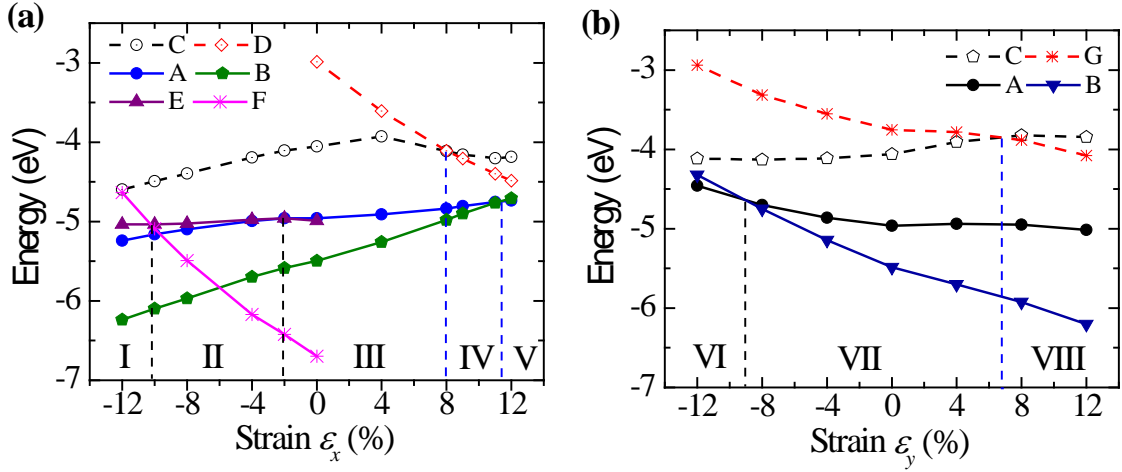

**Figure 6. The energies of the near-edge states A - G as a function of strain applied in the zigzag (left) and armchair (right) direction, respectively. The critical strains for the direct-indirect gap transition are determined by the energy crossover of competing states. In (a), the energy crossover of the competing VB states A, E, and F occurs at $\varepsilon_x$ = -2%, -10.2%, and +11.3% strains. The energies of the competing CB states C and D are equal at $\varepsilon_x$ = +8% strain. In (b), the energies of the VB states A and B crosses at $\varepsilon_y$ = -9%, and two CB states C and G meets at $\varepsilon_y$ = +6.8%. Strain zones I-VIII are also labeled.**

In Figure 6, the conduction states were represented by hollow-dashed lines while valence states given by solid lines. In Figure 6(a), state C represents the CBM from -12% strain up to +8%, at which the energy of state D equals to that of C. From +8% to +12% strain, D has a lower energy than C thus D represents the CBM. The energy crossover of these two states gives the critical strain +8% for the direct-indirect band gap transition.

For the valence bands, two states A and B compete for the VBM at the positive strain, while three states A, E and F battle at compression. State A represents the VBM for positive strain up to +11.3%, where C catches up and becomes the VBM. For the negative strain, the energy of A was first exceeded by E at -2% strain and E becomes the VBM in the strain range from -2% to -10.2%. At -10.2% strain, F catches up and represents the VBM for compression higher than -10.2%.

In the case of strain $\varepsilon_y$ in Figure 6(b), the energy crossover of two conduction states C and G occurs at +6.8% strain and that of two valence states A and B at -9%.

In order to understand the different trends of states A-G in Figure 6, their electronic orbitals, including charge distribution and wavefunction, were explored. The wavefunction character was examined by projecting the wavefunction onto s-, p-, and d-orbitals at each ionic site. It was found that states A and E are $p_z$-orbitals; B is $p_y$; C is dominated by $p_z$ (86%) mixed with $p_y$; State D is dominated by $p_y$ (56%) with a mixture of s-orbital; F is $p_x$; and G is dominated by $p_x$ (59%) mixed with s-orbital. Their electron density contour plots and wavefunction character are presented in Figure 7.

We developed a general mechanism [36] to explain the different energy shifts with strain based on the Heitler-London's exchange energy model [51]. The different energy shift with strain was found to be closely related to the bonding/anti-bonding nature of the orbitals [36]. In the Heitler-London's model, the energies of the bonding and antibonding states are given by the equations,

$$E_{bonding} = 2E_0 + \frac{e^2}{R} + \frac{K+H}{1+S^2} \qquad (1)$$

$$E_{antibonding} = 2E_0 + \frac{e^2}{R} + \frac{K-H}{1-S^2} \qquad (2)$$

where $E_0$ is the energy for an isolated atom, $K$ represents the classical Coulomb energy between the electron-electron and electron-ion interactions, and $S$ is the overlap integral of the orbitals between different atomic sites, which is usually much smaller than 1. The square of $S$ is even smaller. Therefore, the exchange integral term $H$ may be playing a dominant role in determining the different energy variation behaviors with strain in the bonding and antibonding situation. The exchange $H$ is given by,

$$H = \iint \psi_a^*(r_1)\psi_b^*(r_2)(\frac{1}{r_{12}} - \frac{1}{r_{2a}} - \frac{1}{r_{1b}})\psi_b(r_1)\psi_a(r_2)dr_1dr_2 \qquad (3)$$

where the exchange $H$ is contributed from either non-classical electron-electron (i.e. $\frac{1}{r_{12}}$, positive) or electron-ion interaction (i.e. $-\frac{1}{r_{2a}} - \frac{1}{r_{1b}}$, negative). For s-orbitals or any mixed orbitals in which electron density are not extremely localized, the contribution of the electron-ion interaction is dominated over the electron-electron interaction in the exchange $H$. As the atomic distance increases (corresponding to a positive tensile strain), the energy contributed from the electron-ion interaction increases more rapidly compared to the energy reduction of the electron-electron contribution (see Equation (3)), which results in an increased value for $H$. And an increased $H$ value causes the bonding energy $E_{bonding}$ to increase and the antibonding energy $E_{antibonding}$ to decrease based in Equations (1) and (2). These trends are represented by the schematic in Figure 7(h). For the case of non-bonding, in which the wavefunction overlap is minimal, the energy is expected to be insensitive to strain.

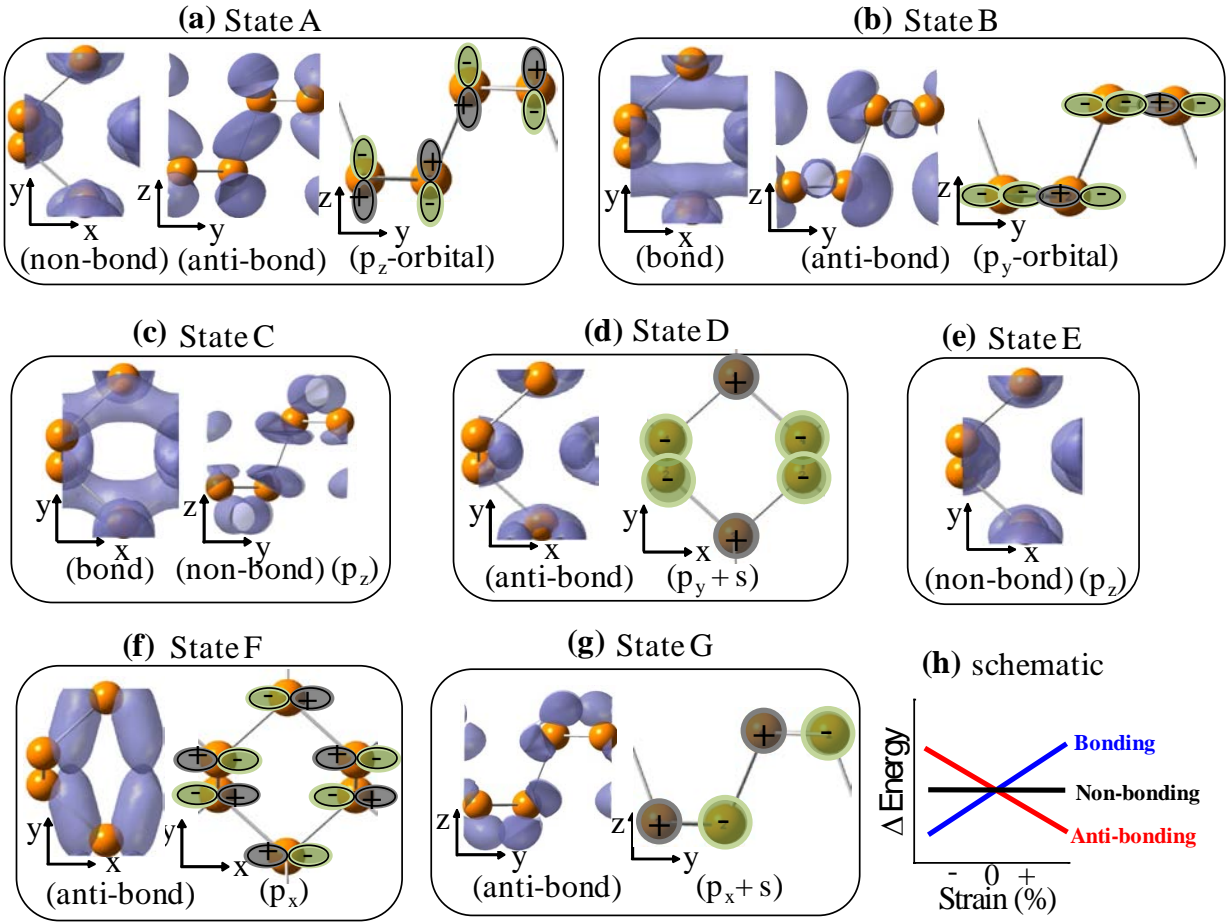

**Figure 7.** **(a)-(g) The electron density contour plots and schematic of the wave function character (i.e. the projected major orbital and sign of phase factor) of the near-band-edge states A - G in 2D phosphorene. Their dominated orbitals and bond status (in the horizontal axis) were listed at the bottom of each state. (h) A schematic of energy response to axial strain for three typical cases of bonding, non-bonding, and anti-bonding.**

Now to examine the electron density contour plots and wavefunction character in Figure 7 to determine their bond nature. State A in Figure 7(a) suggests non-bonding in the x and an anti-bonding character in the y direction based on its sign of phase factor along the y direction. B is bonding in the x while displaying anti-bonding in the y direction. C, which is dominated by $p_z$-orbital, illustrates a bonding nature in the x while non-bonding in the y direction. States D, E and F demonstrate anti-bonding, non-bonding and anti-bonding, respectively, in the x axis. State G is anti-bonding in the y direction. State D is a mixture of $p_y$ and s-orbitals. Since the overlap of the $p_y$ orbital in the x direction is small, the s-orbital is plotted in Figure 7(d) to determine its bonding/anti-bonding status in the x direction. The same case is applied to State G in Figure 7(g).

The bond nature of these seven states combined with the schematic in Figure 7(h) can be used to explain their energy variation in Figure 6. For example, D is anti-bonding in the x direction and expected to decrease with tensile strain from Figure 7(h). This is in agreement with the curve of D in Figure 6(a). B is bonding in the x while anti-bonding in the y direction. According to Figure 7(h), its energy is expected to increase with $\varepsilon_x$ while decrease with $\varepsilon_y$, which are consistent with Figure 6. Other curves in Figure 6 can be explained in the same matter.

**Strain Effect on the Effective Masses of Charge Carriers.** The effective masses of the electron and hole can be readily calculated according to the formula $m* = \hbar^2 (\frac{d^2 E}{dk^2})^{-1}$ from the band structure. For relaxed phosphorene, the effective mass of the electron is predicted to be 1.24 $m_e$ in the zigzag and 0.16 $m_e$ in the armchair direction. The effective mass of the hole is 4.92 $m_e$ in the zigzag and 0.15 $m_e$ in the armchair direction. These calculated effective masses are in agreement with other theoretical work [25, 26]. The significant smaller effective masses in the armchair direction suggest that charge carriers prefer to transport in that direction.

The effective masses of the electron and hole as a function of both strains $\varepsilon_x$ and $\varepsilon_y$ are presented in Figure 8. The effective masses of the electron and hole can be dramatically tuned by strain. In addition, it was found that the sudden jump (drop) in the effective masses occurs around the strain zone boundaries (i.e. critical strains) for the direct-indirect band gap transition. For example, in Figure 8(a), the effective mass of the electron in the zigzag direction is an order of magnitude bigger than that along the armchair direction in Zones I, II, and III. At the zone boundary of III and IV, the effective mass in the zigzag direction drops sharply while in the armchair direction jumps suddenly. This sharp transition suggests the zigzag direction is favored for electron transport [25]. Figure 8(b) shows that the hole prefers to transport in the armchair direction in Zones II-V. However, in Zone I, it favors the zigzag direction.

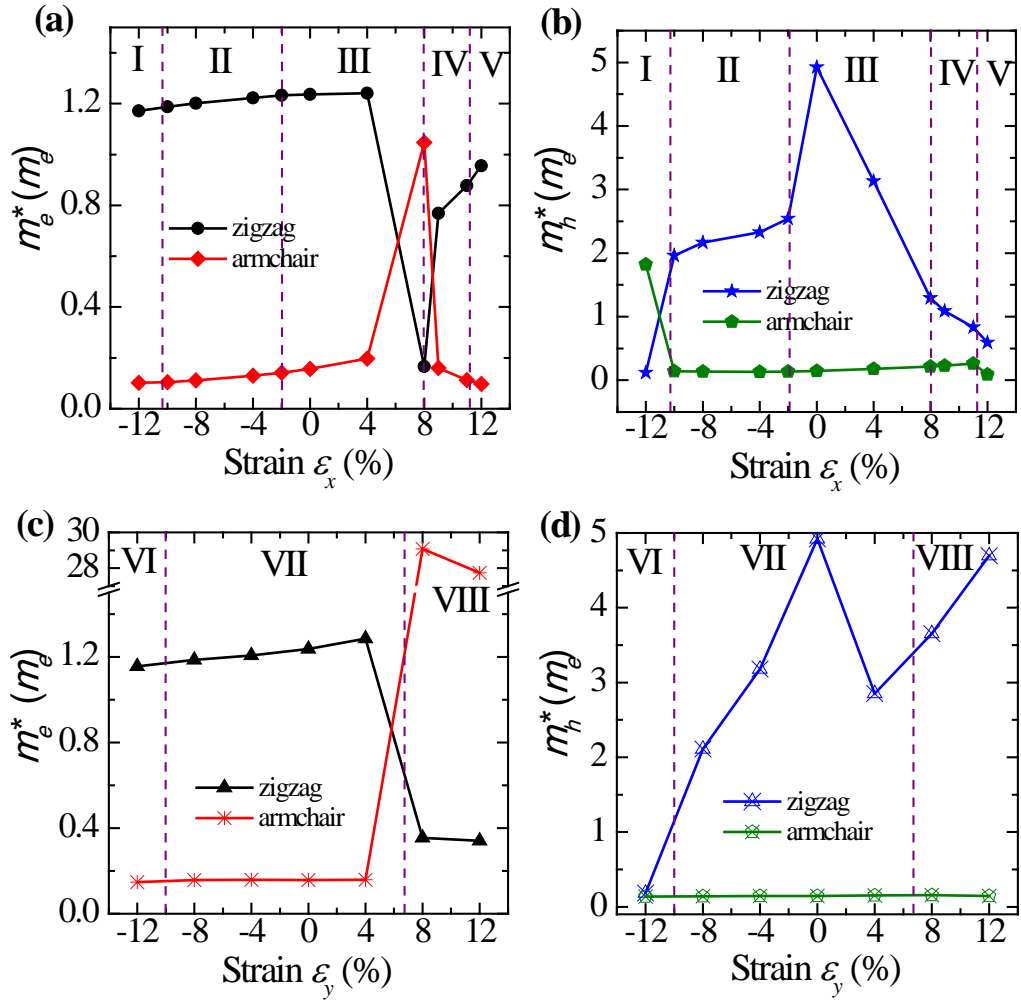

**Figure 8. Effective masses of the electron (left) and hole (right) as a function of strain $\varepsilon_x$ (top) and $\varepsilon_y$ (bottom). Five (three) strain zones for $\varepsilon_x$ ($\varepsilon_y$) are also labeled. The sharp shift in effective masses occurs around the zone boundaries for the direct-indirect gap transition.**

For the case of strain applied in the y direction, Figure 8(c) suggests the armchair direction is favored for the electron transport in Zones VI and VII, while the situation becomes opposite in Zone VIII. For the hole in Figure 8(d), the effective mass in the armchair direction is insensitive to strain, while in the zigzag direction it experiences a dramatic shift. The much smaller value of effective mass in the armchair direction indicates the hole predominately prefers transport in the armchair direction in Zone VII and VIII. However, the sharp reduction of the effective mass in the zigzag direction at Zone VI makes this axis is competitive with the armchair direction for the hole transport. .

The effective mass presented in Figure 8 is a direct consequence of the strain effect on the band structure in figures 3 and 4. In particular, the sharp change in the effective masses results from the direct-indirect band gap transition. For example, to understand the sudden shift of the effective mass at $\varepsilon_x = +8\%$ in Figure 8(a), please refer to the band structure of Figure 3(h) along $\Gamma$ to Y. Compared to +4% strain in Figure 3(g), the downward shift of state D at +8% strain largely reduces the band dispersion at state C, thus increasing the effective mass of the electron dramatically at $\varepsilon_x = +8\%$. When $\varepsilon_x$ is bigger than +8%, the CBM is shifted away from C to D, thus the effective mass of the electron was calculated from D, which has a much smaller effective mass resulting from the more dispersive band structure. Another sharp shift in the effective mass of the hole occurs at $\varepsilon_x = -12\%$ in Figure 8(b), which is a direct consequence of Figure 3(a). At $\varepsilon_x = -12\%$, the energy of sub VB state F exceeds E and becomes the VBM. The effective mass along the x direction is now calculated based on this new state F instead of E. Similarly for the case of $\varepsilon_y$ strain, the striking transition of the effective mass of the electron from +4% to +8% strain in Figure 8(c) is resulted from the +6.8% critical strain.

**Summary**: Using *ab initio* calculations, we provided a detailed analysis of strain effects on the electronic band structure of 2D phosphorene. We found that (1) phosphorene can withstand a surface tension and tensile strain up to 10 N/m and 30%, respectively. (2) The band gap of 2D phosphorene has direct-indirect-direct transitions with axial strain. (3) Five strain zones with distinct electronic band structure were identified and the critical strains for the zone boundaries were determined. (4) The origin of the gap transition was revealed and a general mechanism was developed to explain the near-band-edge energy shifts according to the bond nature of their electronic orbitals. (5) In relaxed phosphorene, effective masses of the electron and hole in the armchair direction are an order of magnitude smaller than that of the zigzag direction suggesting that the armchair direction is favored for carrier transport. (6) Effective masses can be dramatically tuned by strain. (7) The sharp jump/drop in the effective masses occurs at the zone boundaries of direct-indirect gap transition.

Phosphorene has demonstrated superior mechanical flexibility and can hold a large tensile strain of 30%, which opens doors for applications in flexible display. Having a direct band gap is essential for materials in optical applications. Our work shows that a moderate -2% stain in the zigzag direction can trigger the direct-to-indirect band gap transition. Our predicted strain Zones II, IV, VI, and VIII should be avoided for optical applications, due to the indirect band gap in the material within these strain zones. Carrier mobility is an essential parameter to determine performance of electronics and it is inversely dependent on the effective masses. We demonstrated that strain can dramatically tune the effective masses, thus can be used to modify carrier mobility of phosphorene in electronics applications.

## AUTHOR INFORMATION

**Corresponding Author,** * E-mail: xihong.peng@asu.edu.

**ACKNOWLEDGMENT**

This work is supported by the Faculty Research Initiative Fund from School of Letters and Sciences at Arizona State University (ASU) to Peng. The authors thank ASU Advanced Computing Center for providing resources. Drs. F. Tang and L. Yang are acknowledged for the helpful discussions. Dr. Tang is also greatly appreciated for the critical review of the manuscript.